\documentclass[pra,twocolumn,showpacs,preprintnumbers,amsmath,amssymb,superscriptaddress]{revtex4}

\usepackage{graphicx}
\usepackage{dcolumn}
\usepackage{bm}
\begin{document}
\draft
\newcommand{\lw}[1]{\smash{\lower2.ex\hbox{#1}}}

\title{Polarization Plateau in Atomic Fermi Gas Loaded on Triangular
Optical Lattice}

\author{M.~Okumura}
\email{okumura.masahiko@jaea.go.jp}
\affiliation{CCSE, Japan Atomic Energy Agency, 6--9--3 Higashi-Ueno,
Taito-ku, Tokyo 110--0015, Japan}
\affiliation{CREST (JST), 4--1--8 Honcho, Kawaguchi, Saitama 332--0012,
Japan}
\author{S.~Yamada} 
\email{yamada.susumu@jaea.go.jp}
\affiliation{CCSE, Japan Atomic Energy Agency, 6--9--3 Higashi-Ueno,
Taito-ku, Tokyo 110--0015, Japan}
\affiliation{CREST (JST), 4--1--8 Honcho, Kawaguchi, Saitama 332--0012,
Japan}
\author{M.~Machida}
\email{machida.masahiko@jaea.go.jp}
\affiliation{CCSE, Japan Atomic Energy Agency, 6--9--3 Higashi-Ueno,
Taito-ku, Tokyo 110--0015, Japan}
\affiliation{CREST (JST), 4--1--8 Honcho, Kawaguchi, Saitama 332--0012,
Japan}
\author{T.~Sakai}
\email{sakai@spring8.or.jp}
\affiliation{Japan Atomic Energy Agency, SPring-8, Sayo, Hyogo
679--5148, Japan}
\affiliation{Department of Material Science, University of Hyogo,
Kamigori, Hyogo 678--1297, Japan} 

\date{\today}

\begin{abstract} 
 In order to demonstrate that atomic Fermi gas is a good experimental
 reality in studying unsolved problems in frustrated interacting-spin
 systems, we numerically examine the Mott core state emerged by loading 
 two-component atomic Fermi gases on triangular optical lattices. 
 Consequently, we find that plateau like structures are observable
 in the Mott core polarization as a function of the population
 imbalance. These plateau states are caused by a flexibility that the
 surrounding metallic region absorbs the excess imbalance to keep the
 plateau states inside the Mott core. We also find spin patterns
 peculiar to the plateau states inside the Mott core.  
\end{abstract}

\maketitle

Interplay between geometrical frustration and quantum fluctuation in
low-dimensional condensed matters allows non-trivial ground state. One
of the typical example is the ground state of the isotropic triangular
$S=1/2$ antiferromagnetic Heisenberg and {\it XXZ} models. Anderson and
Fazekas proposed that the resonating valence bond state is a candidate
of the ground state \cite{RVB}. Contrary to the conjecture, numerical
calculations suggested 120$^\circ$ N{\' e}el-ordered ground state
\cite{NumTL}. On the other hand, several experiments showed that the
ground state is a spin liquid \cite{He,Et,SLExp1,SLExp2,SLExp3}. But,
its excitation feature differs even in experiments using the same
materials. One claimed gapless \cite{SLExp2}, and another gapful
\cite{SLExp3}. These controversies require more systematic and precise 
experiments. 

Another typical example is 1/3 magnetization plateau state in triangular
antiferromagnetic Heisenberg and {\it XXZ} models under the magnetic
field. Miyashita suggested the existence based on classical analysis
\cite{Miyashita}, and both numerical calculations
\cite{Nishimori,Honecker,Okunishi,Miyahara} and experiments
\cite{1o3Exp} confirmed it. In addition, Oshikawa, Yamanaka, and Affleck
(OYA) advocated a related quantization condition for systems with
periodic boundary condition given by $q ({\cal S} - m) = {\rm integer}$,
where $q$ is an integer being the size ratio of the ground-state
unit-cell to the original one and ${\cal S}$ and $m$ are the spin
quantum number and the average magnetization per unit cell of the
system, respectively \cite{OYA}. The OYA quantization condition predicts
how a translational symmetry is broken and how a magnetization value
causing the plateau is given. In fact, using the density-matrix
renormalization group (DMRG) method \cite{White,DMRGreview}, Okunishi
and Tonegawa actually confirmed a spin distribution, in which a
translational symmetry is broken, in a zigzag chain model
\cite{Okunishi}. Presently, it is known that 1/3 plateau in the {\it
XXZ} model exists in both the quantum model and the classical
counterpart while 1/3 plateau in the Heisenberg model and 2/3 plateau in 
both models arises from pure quantum origin
\cite{Miyashita,Nishimori,Okunishi}. Moreover, there exists a critical
ratio of the distortion of the triangular lattice in the Heisenberg
model and in the {\it XXZ} model for existence of the plateau
\cite{Nishimori,Honecker,Okunishi,Miyahara,Alicea}. Thus, a remaining 
issue is to clear how the plateau states appear and disappear in
parameter variations of frustrated models. This clearly requires a quite 
controllable experiment for the distortion. 

\begin{figure}
 \includegraphics[scale=0.48]{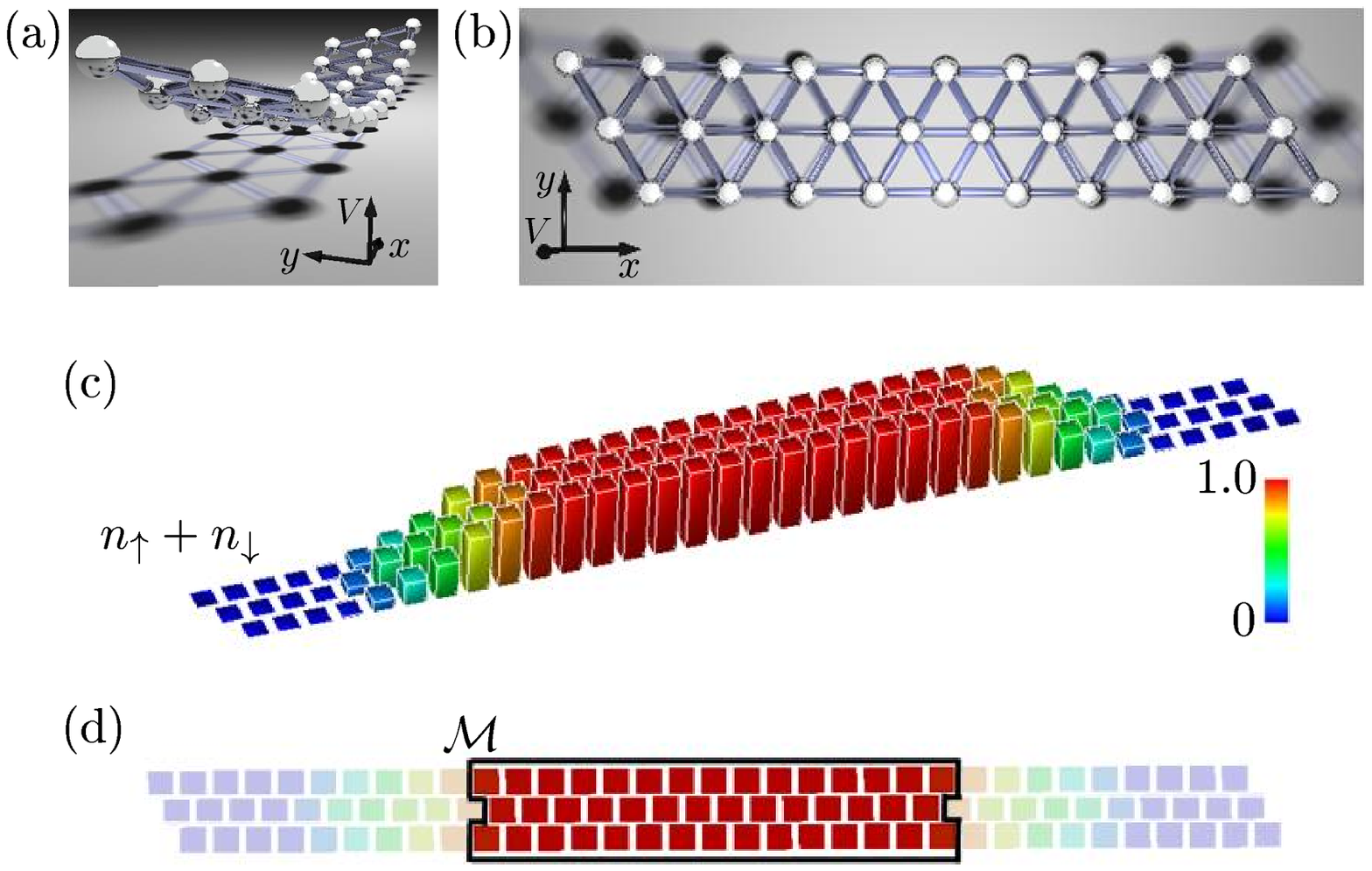}
 \caption{\label{fig1} (a), (b) Schematic figures of the system
  considered in this paper. The $x$- and $y$-directions are taken for
  leg- and rung-direction, respectively. The potential strength is
  displayed in the perpendicular direction to the $x$-$y$ plane. (c)
  A typical density distribution of 60 fermions with $U/t=10$ and
  $V/t=0.07$. The Mott core is formed in the center of the system. The
  atom density ($n_\uparrow + n_\downarrow$) profile does not depend on
  the spin imbalance ratio $p$. (d) The Mott core region ${\mathcal M}$,
  where the density ($n_\uparrow + n_\downarrow$) is a unit.
 }
 \end{figure}

In this paper, we suggest that atomic Fermi gas loaded on optical
lattice (FGOL) \cite{CAreview} can offer a crucial stage on such
controversial topics. Our propose is based on the following features 
peculiar to FGOL. The repulsively interacting two-component FGOL creates
the so-called Mott core in the central region due to the existence of
both a repulsive interaction and a harmonic trap brought about by the
laser intensity profile \cite{Rigol,Machida} as schematically shown in
Fig.~\ref{fig1}(a) and \ref{fig1}(b). The Mott core state, which was
recently confirmed experimentally \cite{MottCoreExp1,MottCoreExp2}, is
mapped onto $S=1/2$ Heisenberg spin model, since the spin degree of
freedom survives solely. In a triangular optical lattice, which is
easily accessible in FGOL, the distortion is almost freely controllable
by adjusting the angle between the counter laser beams, and the
arbitrary anisotropy variation is also achievable by producing
spin-state dependent difference on the atom hopping
\cite{Duan}. Moreover, the application of the magnetic field is mimicked
by changing the population imbalance \cite{MottHeisenberg}. Thus, FGOL
is found to be a quite flexible experimental reality to study quantum
frustrated systems systematically. However, we note that there is only a
big difference between condensed-matter spin systems and FGOL. The Mott
core as stage of spin models is surrounded by a metallic periphery
having a role of ``environment'' to the spin system. Then, it is not
obvious whether or not the feature is insignificant in studying the
above critical issues. Therefore, main aims of this paper are to perform
direct simulations of the trapped frustrated FGOL and to actually
confirm the advantage of use of FGOL. In this paper, we concentrate on
1/3 plateau and related states as a trial problem on the trapped
frustrated FGOL. Other issues are now under investigations. 

In FGOL, the magnetic field and the magnetization are replaced by the
imbalance ratio and the polarization, respectively. We solve a
triangular Hubbard model with a harmonic trap potential for 
$x$-direction [Fig.~\ref{fig1}(a) and \ref{fig1}(b)] given by 
\begin{align}
 H & = -t\sum_{\sigma, \langle {\bm i}, {\bm j} \rangle} c_{\sigma {\bm  
 i}}^\dag c_{\sigma {\bm j}}^{\phantom{\dag}} + V \sum_{\sigma,{\bm i}}
 \left[ \left( {\bm r}_{\bm i}  - {\bm r}_{\rm c} \right) \cdot {\bm
 e}_x \right]^2 n_{\sigma{\bm i}} \nonumber \\
 & \quad {} + U \sum_{{\bm i}} n_{\uparrow {\bm i}} n_{\downarrow {\bm
 i}} \, , 
\end{align}
where $\langle {\bm i}, {\bm j} \rangle$ refers to the nearest
neighbors, $\sigma=\uparrow$ and $\downarrow$, ${\bm r}_{\bm i}$ and
${\bm r}_{\rm c}$ are the position vector of the ${\bm i}$-th site and
the center of the system, ${\bm e}_x$ is a unit vector for $x$-direction,
$t$ is the hopping parameter, $U$ is the on-site repulsion, $V$ is the
harmonic potential strength, $c_{\sigma {\bm i}}^{\phantom \dag}$
($c_{\sigma {\bm i}}^\dag$) is the annihilation- (creation-) operator
and $n_{\sigma {\bm i}} (\equiv c_{\sigma {\bm i}}^\dag c_{\sigma {\bm
i}}^{\phantom \dag})$ is the site density operator. We calculate the 
imbalance ratio $p (\equiv \sum_{\bm i} N_{-,{\bm i}}/N_{+,{\bm i}})$
vs. the normalized polarization on the emergent Mott core, $M=
\sum_{{\bm i} \in \mathcal{M}} N_{-,{\bm i}}/ N_{+,{\bm i}}$ [see
Fig.~\ref{fig1}(c) and \ref{fig1}(d)], where $N_{\pm,{\bm i}} \equiv 
n_{\uparrow{\bm i}} \pm n_{\downarrow {\bm i}}$. The calculations are
made by using DMRG method \cite{White,DMRGreview} on $3\times34$-sites
triangular ladder with 60 
fermions with $U/t=10$ and $V/t=0.07$. Our DMRG is directly extended to
ladder systems by parallelizing the superblock matrix diagonalization. 
See Ref.~\cite{Yamada} for more details of the parallelization and
related techniques. We confirm the precision of DMRG results in short
3-leg ladders by comparing those of the exact diagonalization, and check
a dependence of results on the number of states kept in long ladders to
obtain reliable results. Then, the number of states kept is changed from
300 to 700 according to the convergence check. 700 is enough for every
case. 

\begin{figure}
\includegraphics[scale=0.7]{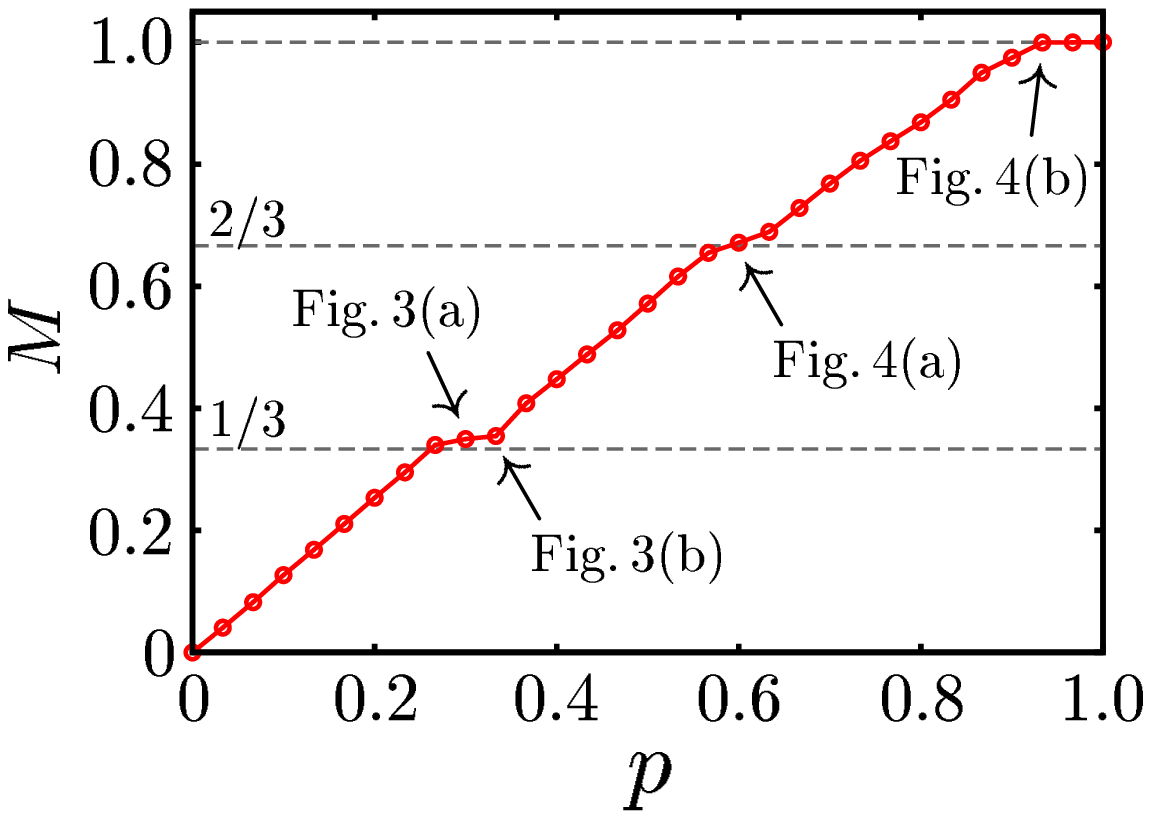}
\caption{\label{fig2} The Mott core polarization ($M$) curve as a
 function of the population imbalance ($p$). The spin distributions at
 the values of $p$ pointed by arrows are given in Figures 3 and 4.
 }
\end{figure}

Let us present DMRG calculation results. Figure \ref{fig2} shows the
spin imbalance ratio $p$ vs. the polarization on the Mott core $M$.
In Fig.~\ref{fig2}, one finds three characteristic features. The first
is a plateau like structure seen around $p=1/3$, the second is a kink
around $p=2/3$, and the third is another plateau one indicating fully
magnetized before $p=1$. According to the OYA quantization condition,
${\mathcal S}=3/2$ is derived in 3-leg ladder. Consequently, we have
$m=1/2$ ($q=1$) and $m=1$ ($q=2$) in 1/3 and 2/3 plateaus,
respectively. These conditions then predict for periodic systems that 
the number of sites in the unit cells of the ground state are given by 
multiples of 3 and 6, respectively in 1/3 and 2/3 plateaus,
respectively. The spin distribution in the 1/3 polarization plateau is 
compared with the OYA prediction in 1/3 magnetization plateau. The 2/3
polarization kink is not a full plateau, but the observed spin structure
is expected to be relevant to the OYA quantization condition in 2/3 
magnetization plateau.

\begin{figure}
\includegraphics[scale=0.62]{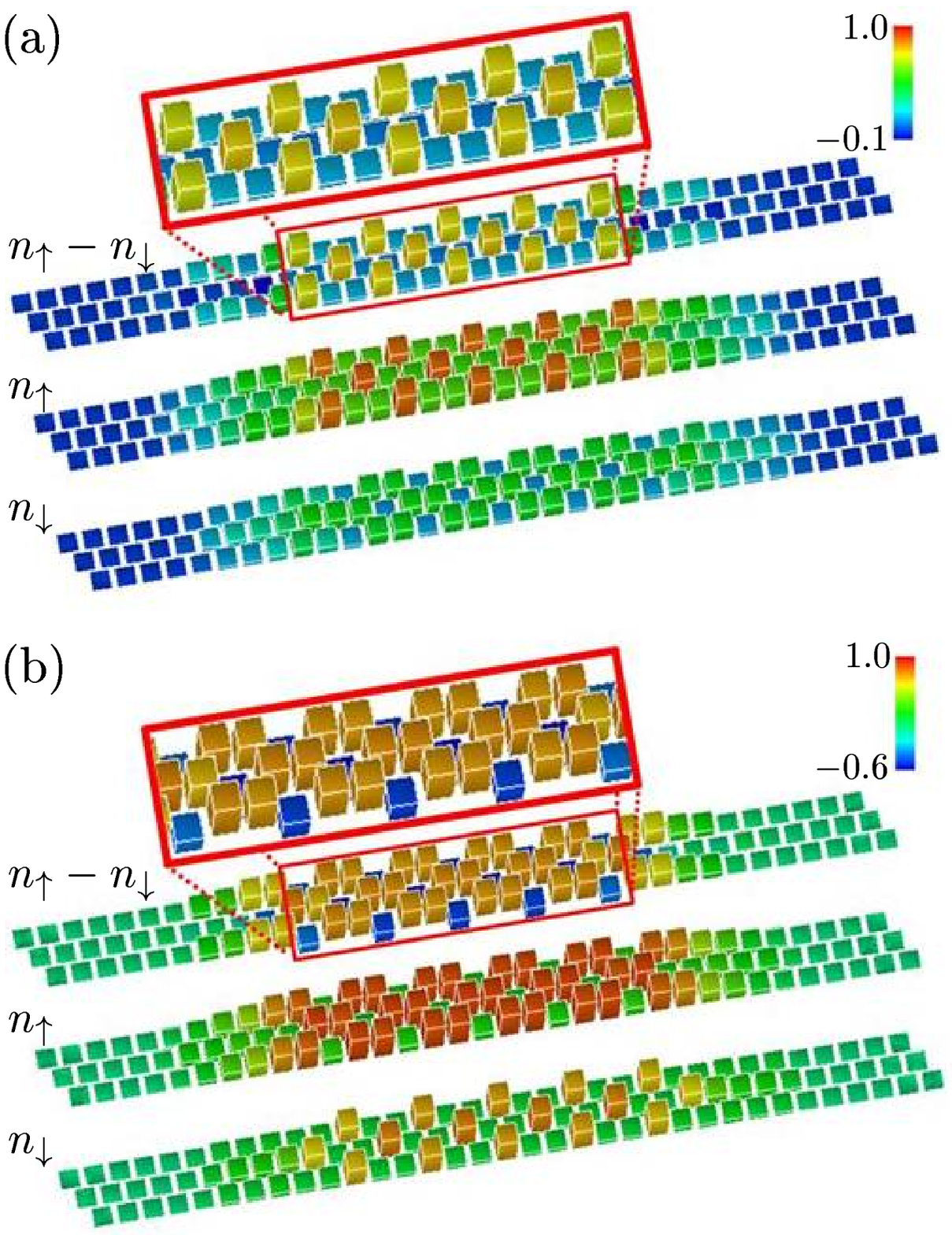}
\caption{\label{fig3} Polarization ($n_\uparrow - n_\downarrow$),
 up-spin ($n_\uparrow$), and down-spin ($n_\downarrow$) density
 distributions at (a) $p=0.30$ and (b) $p=0.33$. The zoomed-in view is 
 the spin distribution on the Mott core.
 }
\end{figure}

Now, let us examine spin distributions on the points indicated in
Fig.~\ref{fig2}. The first focus is 1/3 plateau observed from $p=0.27$ 
to $p=0.33$, in which we find two characteristic patterns. Figure
\ref{fig3}(a) shows a spin distribution pattern at $p=0.30$, and
Fig.~\ref{fig3}(b) another one at $p=0.33$. The pattern at $p=0.27$ is
almost equivalent to Fig.~\ref{fig3}(b) ($p=0.33$). From these patterns,
it is found that both are characterized by the 3-sites periodicity along
the ladder direction. In Fig.~\ref{fig3}(a), {\it up-zero-zero} is
observed along the direction, and {\it up-up-down}
\cite{Miyashita,Nishimori,Honecker,Okunishi,Miyahara,1o3Exp,Alicea} 
in Fig.~\ref{fig3}(b). These results indicate that the polarization 
value is always 1/3 within the Mott core although the population
imbalance ratio is different. It is clearly found that the difference of 
the population imbalance is absorbed in the metallic periphery, i.e.,
``environment'' and 1/3 magnetization is kept inside the Mott core. This 
robustness of the plateau state is crucial for experimental confirmation
in the triangular trapped FGOL. 

\begin{figure}
\includegraphics[scale=0.62]{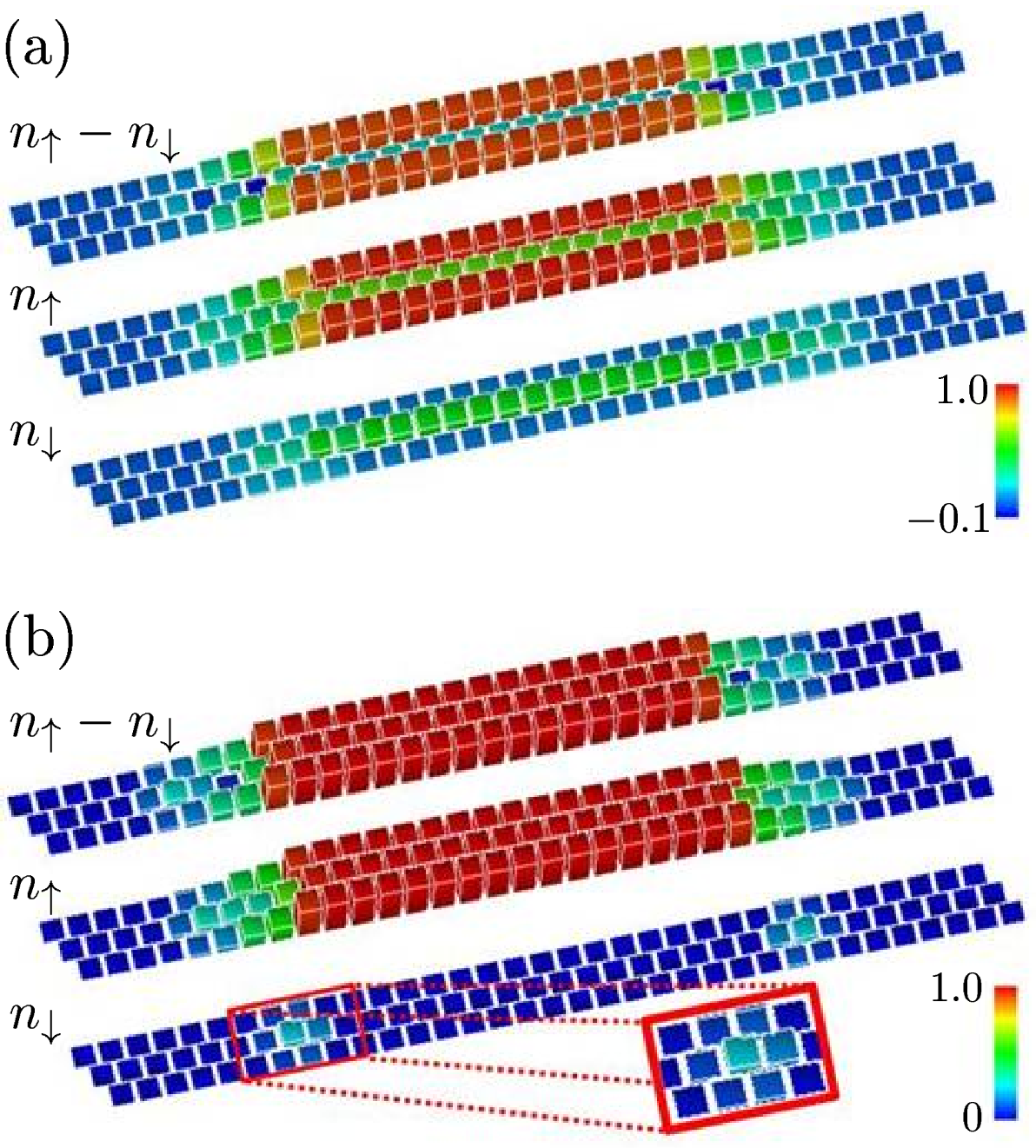}
\caption{\label{fig4} Polarization ($n_\uparrow - n_\downarrow$),
 up-spin ($n_\uparrow$), and down-spin ($n_\downarrow$) density
 distributions at (a) $p=0.60$ and (b) $p=0.93$. The zoomed-in view is 
 the down-spin distribution beside the Mott core.
}
\end{figure}

The next is 2/3 kink observed at $p=0.6$. Figure \ref{fig4}(a) shows the
spin distribution pattern at $p=0.6$. Its key feature is that up-spin and
down-spin species are separately located. The down-spin species almost
assemble along the central line. This brings about an extremely
unbalanced but ordered profiles of the spin density as shown in
Fig.~\ref{fig4}(a). 

The final is the fully-magnetized plateau region before the saturation
imbalance. Figure \ref{fig4}(b) is a typical pattern on the final
plateau. The majority, i.e., up-spin species fully dominate over the
Mott core, while the minority is completely excluded outside the Mott
core. This result also indicates that ``environment'' aids the full
polarization of the Mott core. Such a perfect separation is quite easy
to measure directly. We note that this full separation in addition to
the plateaus is never observed in 3-leg square lattices. 

\begin{figure}
\includegraphics[scale=0.5]{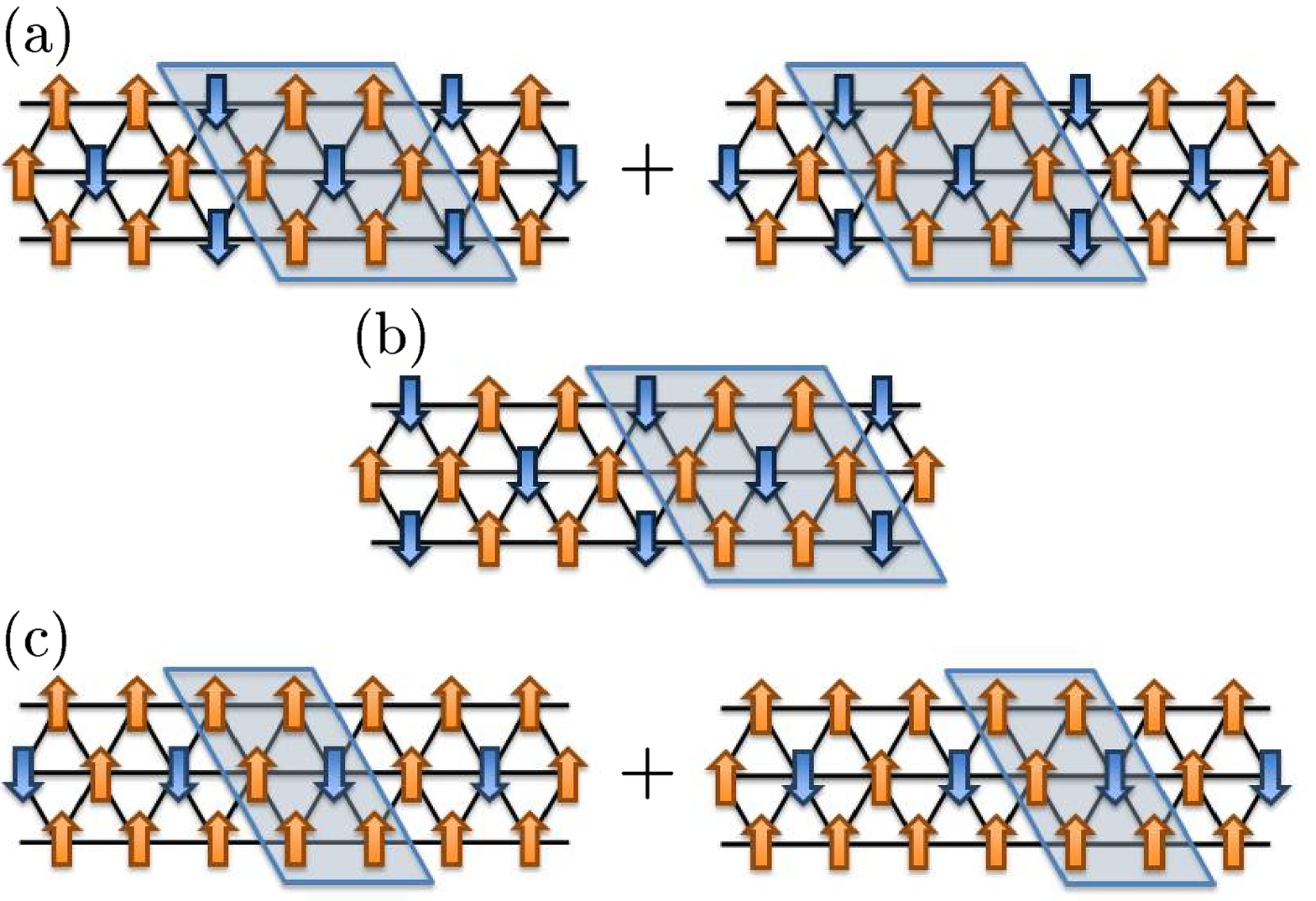}
\caption{\label{fig5} Schematic figures of spin configurations to
 interpret the spin distribution of (a) Fig.~\ref{fig3}(a), (b)
 Fig.~\ref{fig3}(b), and (c) Fig.~\ref{fig4}(a). The shaded area in each 
 figure indicates the unit cell if the system is periodic.
}
\end{figure}

Here, let us discuss why such non-trivial $p$-dependent profiles appear
in the present system. The profile at 1/3 plateau is explained 
as follows. According to the OYA quantization condition, one then
expects that the unit cell of the ground state are given in multiple of 
3 sites at the imbalance range. The profiles observed as
Fig.~\ref{fig3}(a) and \ref{fig3}(b) are consistent with this OYA
prediction and the well-known {\it up-up-down} structure
\cite{Miyashita,Nishimori,Honecker,Okunishi,Miyahara,1o3Exp,Alicea} in
1/3 plateau. The possible profiles and the unit cells with {\it
up-up-down} structure are three patterns given by the right and the left
hand panels in Fig.~\ref{fig5}(a) and one of Fig.~\ref{fig5}(b). Among
them, a consideration about $x$- and $y$-axis inversion symmetries in
the present system leads to a superposition of two profiles as 
Fig.~\ref{fig5}(a) or a profile as Fig.~\ref{fig5}(b). In fact, the
superposition state as Fig.~\ref{fig5}(a) is observed in
Fig.~\ref{fig3}(a), and the state as Fig.~\ref{fig5}(b) coincides with
Fig.~\ref{fig3}(b). In addition, the spin density profile in the 2/3 
polarization kink is relevant to the OYA condition in 2/3 magnetization
plateau. Although observed spin structure in Fig.~\ref{fig4}(a)
indicates that the unit cell of the ground state contains 3 sites, the
structure is reconstructed by a superposition of two states with  
6-sites periodicity predicted by the OYA condition as shown in
Fig.~\ref{fig5}(c). This implies that the observed kink is a shrunk
piece of the 2/3 plateau. The idea can be confirmed by examining a 
situation mapped onto the {\it XXZ} model exhibiting 2/3 plateau
\cite{Nishimori}. The {\it XXZ} model is easily accessible through a
setup of the spin-dependent hopping \cite{Duan}. 

In conclusion, we explored whether or not FGOL is a good stage to study
frustrated quantum spin systems. By applying DMRG method on 3-leg
triangular Hubbard ladder with harmonic potential, we successfully
found 1/3 plateau, 2/3 kink, and full polarization plateau in the Mott
core polarization as a function of the population imbalance. In these
states, we observed that the surrounded metallic region flexibly adjusts
the spin imbalance to keep the plateau states on the Mott core. 
Especially, the minority is completely expelled from the Mott core in
the full polarization plateau. In addition, we point out that the
observed states in 1/3 plateau and 2/3 kink are consistent with the OYA 
predictions. We mention that FGOL can provide a new pathway to study the
quantum frustrated systems. In addition, we point out that the excellent 
controllability is of a great advantage to quantum information
processing using quantum frustrated systems \cite{Pachos}. 

Two of authors (M.O. and M.M.) wish to thank K.~Okunishi and H.~Onishi
for illuminating discussion. The work was partially supported by
Grant-in-Aid for Scientific Research on Priority Area ``Physics of new
quantum phases in superclean materials'' (Grant Nos.~20029019 and
20029020) and for Scientific Research (Grant Nos.~20500044 and
20340096) from the Ministry of Education, Culture, Sports, Science and
Technology of Japan. One of authors (M.M.) is supported by JSPS
Core-to-Core Program-Strategic Research Networks, ``Nanoscience and
Engineering in Superconductivity''. 


\end{document}